\documentclass[a4paper,11pt]{article}
\pdfoutput=1 

\usepackage{jheppub} 

\usepackage[T1]{fontenc} 
\usepackage[english]{babel}
\usepackage{ulem}
\usepackage{xcolor}
\usepackage{amsmath}
\usepackage{tensor}
\usepackage[utf8]{inputenc}
\usepackage{cleveref}
\usepackage{tabularx}

\newcommand{\nc}{\newcommand}
\nc{\cqq}{C_{qq}^{(1)}}
\nc{\cqu}{C_{qu}^{(8)}}
\nc{\theoerror}{\Delta_\text{theo}}
\nc{\mjj}{m_{jj}}
\nc{\pT}{p_\mathrm{T}}
\nc{\lumi}{10\text{\,fb}^{-1}}
\nc{\lumiSeven}{0.7\text{\,fb}^{-1}}

\title{Dijets at Tevatron Cannot Constrain SMEFT Four-Quark Operators}
\preprint{MITP/19-054}
\author[a]{Eduard Keilmann}
\author[b]{and William Shepherd}

\emailAdd{ekeilman@students.uni-mainz.de}
\emailAdd{shepherd@shsu.edu}

\affiliation[a]{PRISMA Cluster of Excellence \& Mainz Institute of Theoretical Physics, \\ Johannes Gutenberg-Universit\"at Mainz, 55099 Mainz, Germany}
\affiliation[b]{Physics Department, Sam Houston State University, Huntsville TX 77341, USA}

\abstract{
We explore the sensitivity of Tevatron data to heavy new physics effects in differential dijet production rates using the SMEFT in light of the fact that consistent and conservative constraints from the LHC cannot cover relatively low cutoff scales in the EFT. In contrast to the results quoted by the experimental collaborations and other groups, we find that, once consistency of the perturbation expansion is enforced and reasonable estimates of theoretical errors induced by the SMEFT series in $\frac{E}{\Lambda}$ are included, there is no potential to constrain four-quark contact interactions using Tevatron data. This shows the general difficulty of constraining physics model-independently using fairly imprecise measurements, limited by low luminosity and/or systematic errors inherent to the precision of the detectors.
}

\begin{document}

\maketitle

\section{Introduction}

Particle physics is firmly entrenched in a new era of data-driven science. The LHC's performance has been spectacularly validated with the discovery of the Higgs boson \cite{Aad:2012tfa,Chatrchyan:2012xdj} and the impressive amounts of data taken at unprecedented speed. This impressive triumph heralds a challenge for our field to determine how best to utilize all this data, which will enable extremely precise measurements of the particles of the Standard Model (SM).

Unfortunately, it seems likely at this point that we will only be able to measure the properties of SM particles, and not the slew of new particles that were widely expected in LHC data. Our current bounds on our favorite models like supersymmetry or extra dimensions are now comparable to the long-term discovery potential of the LHC science program, indicating that we aren't likely to find those particles without a further significant increase in collision energy. It remains entirely possible that there are new particles already in the LHC data which simply don't have signatures similar to those that we initially expected, but it is also quite plausible that whatever new physics (NP) exists to address the remaining questions left open by the SM could be at energy scales the LHC will never be able to probe.

Taking the assumption that new physics is too heavy for the LHC to probe directly, it behooves us to do what we can to learn about its interactions indirectly, and to be as agnostic as possible about the nature of the physics we are searching for. Both of these aims are excellently achieved by utilizing the toolkit of Effective Field Theory (EFT). Using EFT techniques we can exploit the difference in scales between the SM and NP to make predictions about what impacts NP can have on the scattering of SM states, expanded in the scattering energy as a fraction of the characteristic mass scale of NP $\Lambda$.

The particular EFT which we will employ goes by the name of SMEFT, and assumes that the $h\left({\rm 125}\right)$ is in fact the Higgs boson, the remnant of the electroweak doublet whose vacuum expectation value (VEV) $v$ is responsible for the breaking of electroweak symmetry in the vacuum, a presumption well-motivated by the proximity of the latest masurements to the SM Higgs predictions~\cite{ATLAS:2018doi,Sirunyan:2018koj}. With this assumption, it is possible to enforce the full SM $SU(3)_C\times SU(2)_L\times U(1)_Y$ gauge symmetry on the operators of the SMEFT, allowing for a single expansion in the NP scale. For a recent review of these EFT techniques as applied to the SM, see~\cite{Brivio:2017vri}.

Much foundational work has gone into the SMEFT, enabling the sorts of analyses presented here. A complete basis of operators~\cite{Grzadkowski:2010es} and their renormalization behavior~\cite{Jenkins:2013zja,Jenkins:2013wua,Alonso:2013hga} are known, and significant effort has been invested in precision electroweak studies~\cite{Han:2004az,Berthier:2015oma,Berthier:2015gja,Bjorn:2016zlr,Berthier:2016tkq,Ellis:2018gqa,Almeida:2018cld} and loop calculations of observables relevant to both electroweak data and hadron collider searches~\cite{Zhang:2013xya,Gauld:2015lmb,Hartmann:2015oia,Gauld:2016kuu,Maltoni:2016yxb,Zhang:2016omx,Hartmann:2016pil,Baglio:2017bfe,Dawson:2018pyl,Dawson:2018jlg,Vryonidou:2018eyv,Dawson:2018liq,Cullen:2019nnr,Dawson:2018dxp}. The ultimate goal of all this effort is a global fit of all the data in terms of NP scale and Wilson coefficients, with which it will be possible to test any UV model, including those that haven't yet been invented, against the precise measurements made at the LHC and earlier experiments.

Searches for the effects of contact interactions, primarily inspired by models in which the SM particles (most notably quarks) are composite at a high energy scale, have been pursued since the days of the Super Proton Synchrotron collider~\cite{Eichten:1983hw}, and the superficial similarities between the parametrization of these models and a subset of the operators in the SMEFT have inspired the adoption of these searches as bounds on the SMEFT parameter space. However, these searches have always been constructed in such a way that they have made very specific assumptions about the UV structure of the model, and as a result are not directly adoptable as constraints on the full SMEFT, which has been explicitly constructed to avoid the need to make such assumptions.

A consistent and conservative technique to derive bounds on the SMEFT from collider searches has been developed in~\cite{Alte:2017pme} and further applied in~\cite{Alte:2018xgc}; it utilizes the well-understood techniques of perturbation theory in the context of the expansion in inverse powers of the new physics scale $\Lambda$. There are two important steps which, prior to the introduction of this technique, have not been consistently performed in searching for higher-dimension effects in the SMEFT or any other EFT-based framework for model-independent new physics parameterization in collider searches. The first is to consistently apply the perturbation theory expansion and truncation at the level of the measured observable (in this case the dijet production cross section), not an intermediate stage (the amplitude). This is manifestly essential in the case of higher-order calculations which exhibit IR divergences that only cancel when considering both higher-loop and higher-leg interactions together, but is also the generally correct order-by-order calculational technique of perturbation theory in general. The second important step is to make a good-faith estimate of the uncertainty in the calculation that results from higher-order corrections that are neglected in the performed calculation; in the context of renormalizable QFT calculations this is usually accomplished by investigating the renormalization-scale dependence of the result within a fairly arbitrary range, usually taken to be a factor of 2 on either side of the central value.

In this article we utilize the prescription of \cite{Alte:2017pme}, and follow up on the insight of \cite{Alte:2018xgc} that lower-energy colliders can provide constraints complementary to those which can be extracted from the LHC. We thus explore the application of Tevatron data in the dijet channel to the SMEFT. We recast a measurement of the dijet cross section differential in the dijet invariant mass and the pseudorapidity of the forward-most jet~\cite{MandyRominsky:2009yya}. After properly truncating the cross section at $\mathcal{O}\left(\Lambda^{-2}\right)$ and introducing an error term corresponding to the uncalculated effects at $\mathcal{O}\left(\Lambda^{-4}\right)$ we find that there is no region of the SMEFT parameter space to which the Tevatron data is sensitive. We attempt to forecast what improvements could be possible were the full Tevatron dataset to be used to make a comparable measurement, and find that even under the most optimistic assumptions it would remain impossible to consistently constrain the SMEFT utilizing 10 fb$^{-1}$ of Tevatron dijet data.

This article is organized as follows: in the next section, we will quickly review the effect of the SMEFT on dijet production, including our treatments of the signal and theoretical error calculations. In \cref{sec:tevdijet} we will recap the data we used and the forecast assumptions we explored for how it would develop with increasing integrated luminosity, and we will present the analysis techniques and sensitivity measures we were able to achieve in \cref{sec:results}. We will make closing remarks in \cref{sec:conc}.

\section{Dijets in SMEFT}

The Lagrangian of the SMEFT treats the Standard Model interactions as the leading order term in an expansion and supplements it by higher-dimensional operators suppressed by a large mass scale $\Lambda$. All symmetries obeyed by the Standard Model also hold true in the SMEFT. The degrees of freedom are the fields of the SM, and the SMEFT assumes that the Higgs field is the only source of Electroweak symmetry breaking. With this assumption, the SMEFT is invariant under the full $SU(3)_{C} \times SU(2)_{L}\times U(1)_Y$ SM gauge group.

The SMEFT Lagrangian, $\mathcal{L}_\text{SMEFT}$, has the general form
\begin{align}\label{eq:generalSMEFTLagrangian}
\mathcal{L}_\text{SMEFT} &= \mathcal{L}_\text{SM}^\prime + \mathcal{L}^{(5)} + \mathcal{L}^{(6)} + \mathcal{L}^{(7)} + \mathcal{L}^{(8)} + \ldots
\end{align}
where $\mathcal{L}_\text{SM}^\prime$ has the form of the SM Lagrangian. However, higher dimensional operators with Higgs vev insertions correct the would-be SM couplings. The leading effects of this type arise at the order of $v^2/\Lambda^2$. Operators  with $d > 4$ indicate the Lagrangian terms $\mathcal{L}^{(d)}$ composed of higher dimensional operators with dimension $d$; note that this series formally continues to arbitrarily high $d$. The operators at a given dimension can be written as

\begin{equation}\label{eq:SMEFTTerms}
\mathcal{L}^{(d)} = \sum_{k=1}^{N_d} \frac{C_k^{(d)}}{\Lambda^{d-4}} Q_k^{(d)}~,
\end{equation}
where $N_d$ denotes the number of non-redundant operators at dimension $d$, $C_k^{(d)}$ the Wilson coefficients, $Q_k^{(d)}$ the operators which form a basis for corrections at dimension $d$ (we will use the Warsaw basis~\cite{Grzadkowski:2010es}), and the scale at which new degrees of freedom appear due to NP is $\Lambda$.

In order to have an effect that could be observable at colliders, we must select interactions which are not strongly constrained already by lower-energy precision measurements. Therefore, for this analysis of SMEFT effects in dijet production, we consider only $\mathcal{CP}$-even as well as baryon- and lepton-number conserving operators. We also insist that the operators obey the SM $\mathcal{SU}(3)^5$ flavor symmetry, which would be exact in the absence of Yukawa couplings; this assumption is a particularly strong case of Minimal Flavor Violation~\cite{DAmbrosio:2002vsn}.

We focus on operators of dimension-six as these provide the leading contributions in the power series in $\Lambda^{-1}$ to dijet production. At dimension five, only one operator exists, which violates the lepton number $L$ and additionally the $B-L$ symmetry, and famously gives a Majorana mass to the left-handed neutrinos~\cite{Weinberg:1979sa}. A complete set of dimension-seven operators is known, in addition to the Warsaw basis at dimension-six~\cite{Lehman:2014jma}; none of those operators contribute to dijet production. 
In fact, all operators of odd dimensionality have been proven to be lepton-number violating~\cite{Kobach:2016ami}. The operators which could in principle contribute to the production of dijets at leading order are listed in Table~\ref{tab:operatorsOnDijet}. Here, $q$ indicate the left-handed quark doublets, whereas $u$ and $d$ are the right-handed up- and down-type quarks, respectively. The generation of the quarks is encoded in the indices $p,r,s,t$. We denote the generators of the $\mathcal{SU}(3)_C$ gauge group as  $T^A$ with $A\in \{1,\ldots,8\}$ and the generators of  $\mathcal{SU}(2)_L$ as $\tau^I =\sigma^I/2$, where $\sigma^I$ is a Pauli matrix and $I\in \{1,2,3\}$.
\begin{table}[htp]
	\caption{Dimension-six operators of the Warsaw basis contributing to dijet production. Baryon number, lepton number and $\mathcal{CP}$ are conserved by these operators. The operators labelled with an asterisk do not interfere with the QCD amplitude.}
	\footnotesize
	\begin{flushleft}
		\renewcommand{\arraystretch}{1.7}
		\begin{tabularx}{\textwidth}{|c|c||c|c|}
			\hline
			$Q_{qq}^{(1)}$~~~\,  & $(\bar q_p \gamma_\mu q_r)(\bar q_s \gamma^\mu q_t)$& $Q_{qq}^{(3)}$  & $(\bar q_p \gamma_\mu \tau^I q_r)(\bar q_s \gamma^\mu \tau^I q_t)$  \\
			$Q_{uu}$~~~\,      & $(\bar u_p \gamma_\mu u_r)(\bar u_s \gamma^\mu u_t)$ & $Q_{dd}$  & $(\bar d_p \gamma_\mu d_r)(\bar d_s \gamma^\mu d_t)$  \\
			$Q_{ud}^{(1)}$~~$^\ast$  & $(\bar u_p \gamma_\mu u_r)(\bar d_s \gamma^\mu d_t)$ & $Q_{ud}^{(8)}$   &  $(\bar u_p \gamma_\mu T^A u_r)(\bar d_s \gamma^\mu T^A d_t)$ \\
			$Q_{qu}^{(1)}$~~$^\ast$    & $(\bar q_p \gamma_\mu q_r)(\bar u_s \gamma^\mu u_t)$ &
			$Q_{qu}^{(8)}$         &  $(\bar q_p \gamma_\mu T^A q_r)(\bar u_s \gamma^\mu T^A u_t)$ \\ 
			$Q_{qd}^{(1)}$~~$^\ast$ & $(\bar q_p \gamma_\mu q_r)(\bar d_s \gamma^\mu d_t)$ 
			& $Q_{qd}^{(8)}$ &  $(\bar q_p \gamma_\mu T^A q_r)(\bar d_s \gamma^\mu T^A d_t)$ \\
			$Q_G$~~$^\ast$& \rule{45.16pt}{0pt} $f^{ABC} G_\mu^{A\nu} G_\nu^{B\rho} G_\rho^{C\mu} 	\hspace{45.16pt}$&& $\rule{163.235pt}{0pt}$\\ 
			\hline 
		\end{tabularx}
	\end{flushleft}
	\label{tab:operatorsOnDijet}
\end{table}%

The leading-order effect of SMEFT operators on dijet production is the interference of the higher-dimensional operator amplitude with the QCD amplitude. The contribution of dimension-six operators squared are of the same order as the interference terms of dimension-eight operators with the SM amplitude. Given that we do not have a basis of operators at dimension-eight, there are unknown corrections of this order to the cross section which should be treated as theoretical errors, much akin to the estimation of higher-order effects in QCD or QED perturbation theory. Thus, the cross section has this form:
\begin{equation}\label{eq:consistency}
\sigma\propto|\mathcal{A}|^2 = \underbrace{|\mathcal{A}_\text{SM}|^2}_{\text{SM background}} +
\underbrace{\frac{2C_6}{\Lambda^2}\, \mathrm{Re}(\mathcal{A}_\text{d6}\mathcal{A}^\ast_\text{SM})}_{\text{signal}} + \underbrace{\frac{C_6^2}{\Lambda^4}|\mathcal{A}_\text{d6}|^2 + \frac{2C_8}{\Lambda^4}\, \mathrm{Re}(\mathcal{A}_\text{d8} \mathcal{A}^\ast_\text{SM})}_{\text{theoretical uncertainty}} + \ldots
\end{equation}

The interference term is manifestly a linear function of the Wilson coefficients of the operators appearing in \cref{tab:operatorsOnDijet}. However, some of these operators actually do not contribute to the interference term. $Q_G$ does not contribute because it only couples combinations of gluons which have a total helicity which differs from that required by the SM couplings~\cite{Azatov:2016sqh}, and the remaining starred operators similarly link fields in ways which QCD is unable to. As an example, $Q_{ud}^{(1)}$ insists that the up and anti-up quarks form a color singlet, which means that they would not be able to couple to a gluon, and the difference in flavor between the up and down quark bilinears prevents QCD from coupling to a color-octet combination of quarks across the two different bilinears in the operator, as it does for instance in the case of the operator $Q_{dd}$ to yield an interference contribution to the dijet cross section.

\begin{figure}[t]
\includegraphics[width=.45\textwidth]{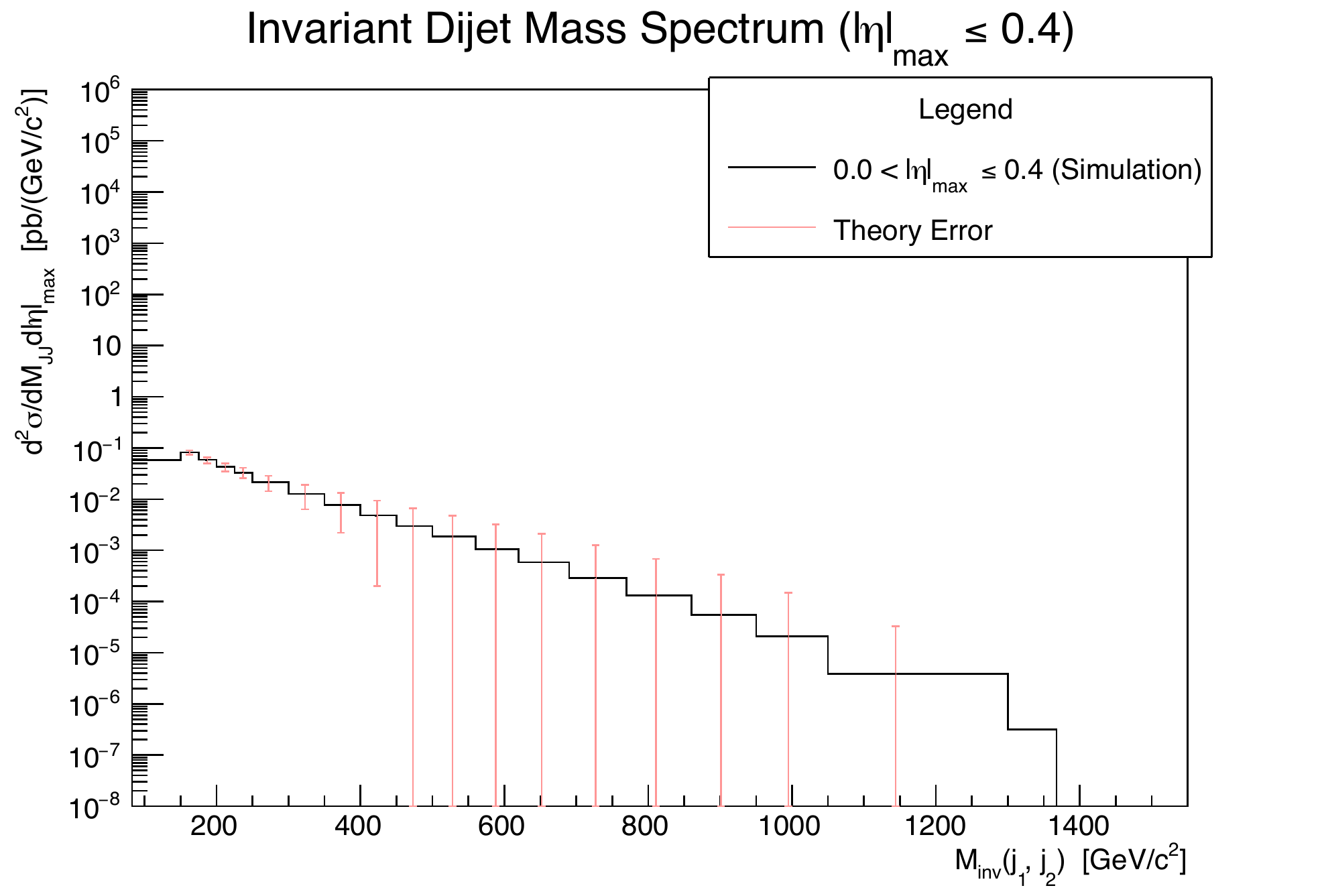}\hspace{0.1\textwidth}
\includegraphics[width=.45\textwidth]{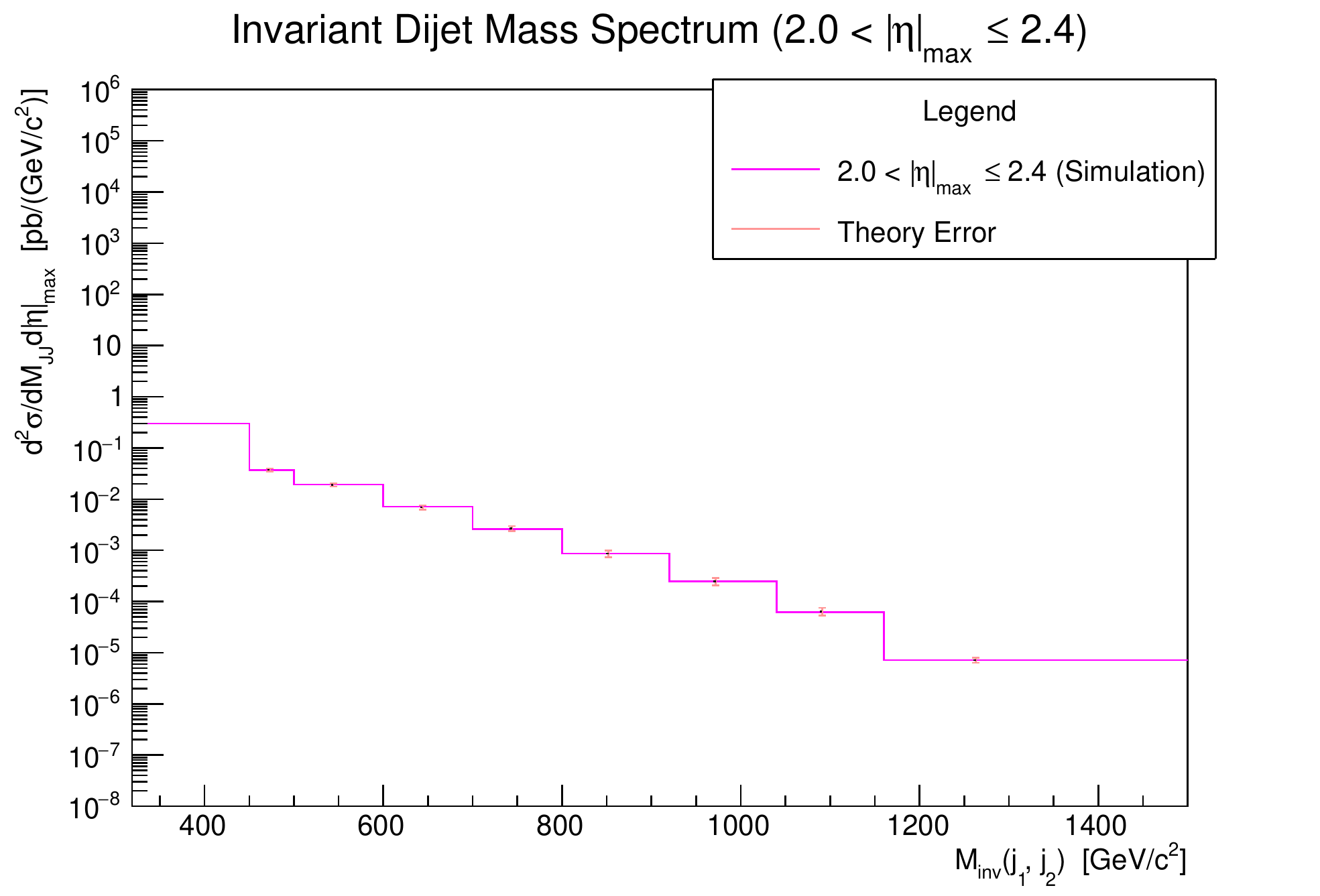}\\
\includegraphics[width=.45\textwidth]{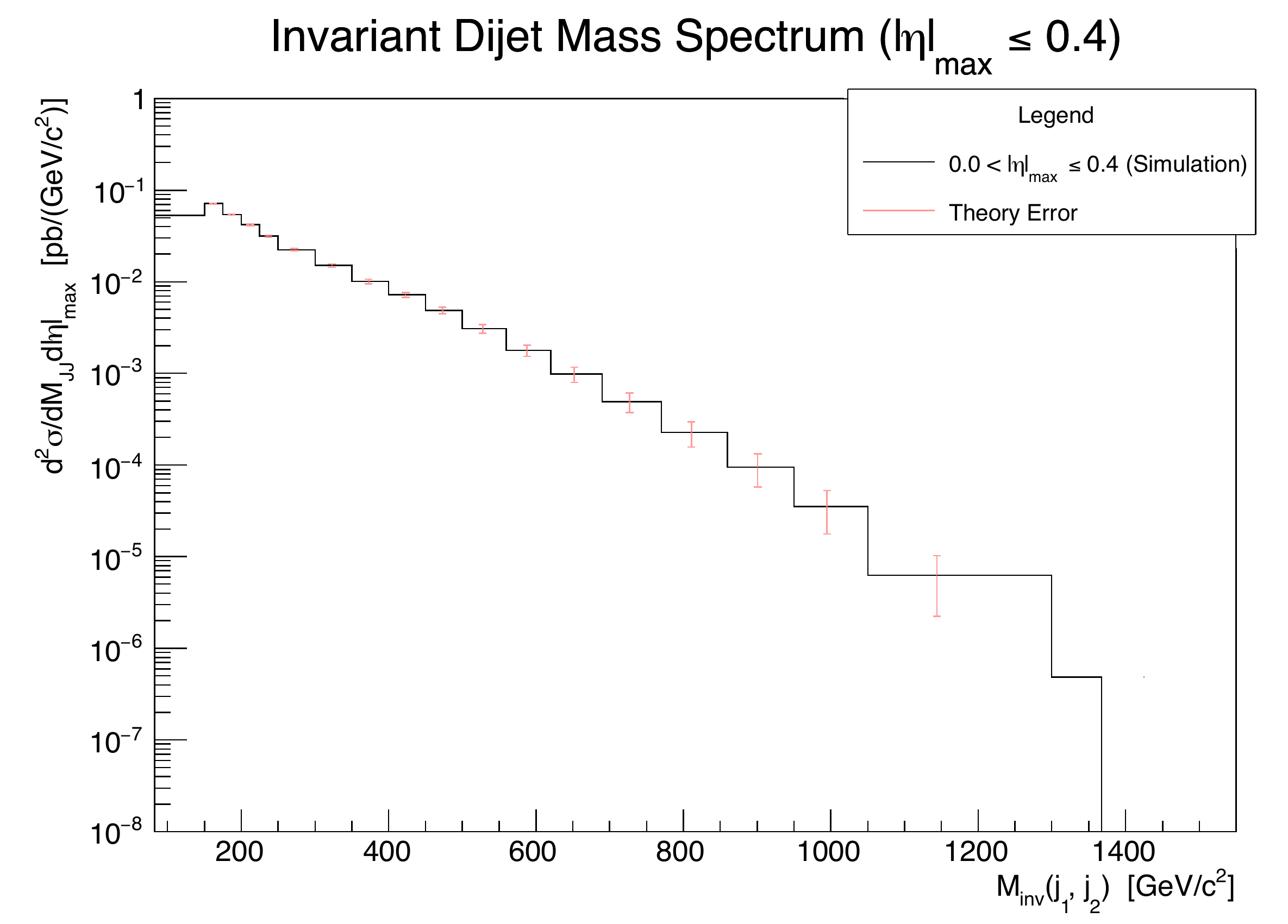}\hspace{0.1\textwidth}
\includegraphics[width=.45\textwidth]{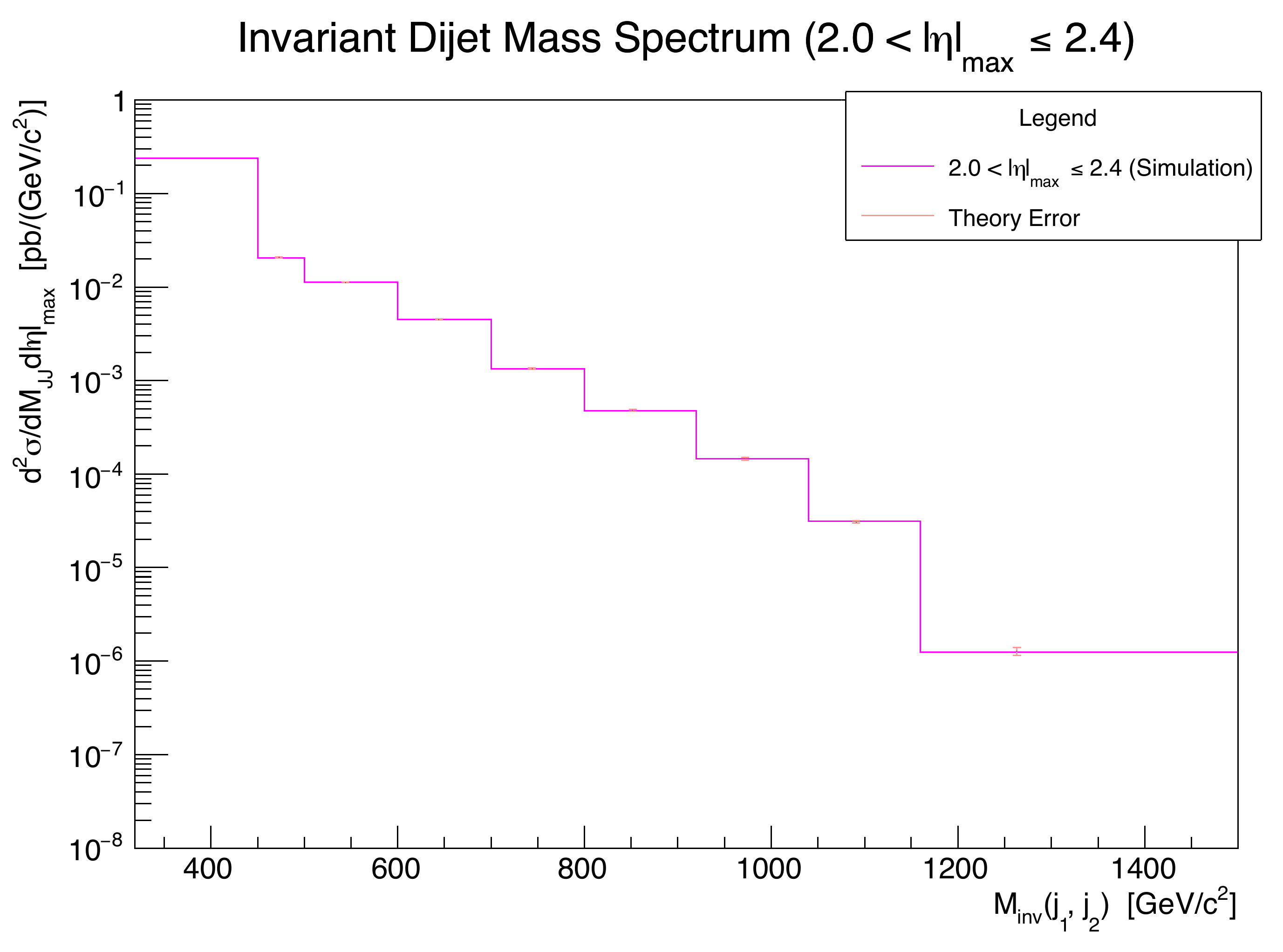}
\caption{\label{fig:sigdist}Signal distributions with theory errors indicated with error bars due to the operator $\cqq$ (upper row) and $\cqu$ (lower row), generated for a Wilson coefficient of -1 and a cutoff scale $\Lambda=1$ TeV. The most central (left column) and most forward (right column) $|\eta|_{\rm max}$ bins are displayed for each operator.}

\end{figure}

We find that there are actually only two different combinations of Wilson coefficients which give measurably different distributions at a hadron collider; one of these produces more central (lower-rapidity) events, and the other produces more forward events. The relevant linear combinations are

\begin{equation}\label{eq:centralDistr}
\frac{\mathrm{d}\sigma}{\mathrm{d}\chi}\bigg \vert_{\text{central}} \propto -(\cqq  + 0.85 C_{uu} + 0.61 C_{qq}^{(3)} + 0.20 C_{ud}^{(8)} + 0.15 C_{dd}  ),
\end{equation}

\begin{equation}\label{eq:flatDistr}
\frac{\mathrm{d}\sigma}{\mathrm{d}\chi}\bigg \vert_{\text{flat}} \propto - (\cqu + 0.45 C_{qd}^{(8)}).
\end{equation}
The coefficients defining these linear combinations depend slightly on partonic collision energy due to differences in PDF prevalences of the different quark species, but these variations are too small to expect to be able to pick apart the different operator's contributions to the observable. The distributions generated by these linear combinations are shown in \cref{fig:sigdist}; we have chosen to turn on the operator with greatest contribution to each of these linear combinations to generate these distributions, and will use these operators as exemplars of the effects of the full linear combinations; note that this choice is easily mapped back to the full general case by using \cref{eq:centralDistr,eq:flatDistr}.

Since the contribution of dimension-six operators squared are of the same order as the unknown interference terms of dimension-eight operators with the SM ones, we utilize the dimension-six squared effect as a proxy for the generic size of this term. We thus consider ourselves to have an error of the form
\begin{equation}
\sigma_{\rm err}\propto \Delta_{\rm theo} |\mathcal{A}_\text{d6}|^2,
\end{equation}
where
\begin{equation}\label{eq:theoError}
\Delta_\text{theo} := \sqrt{C_k^4 + \left(g_s^2 \, c_8 \sqrt{N_8}\right)^2}~,
\end{equation}
has been chosen to incorporate the effects of both the dimension-six squared and the unknown dimension-eight operators, where $g_s$ is the strong coupling of the SM. We are forced to estimate the characteristic size of the dimension-eight Wilson coefficients $c_8$ as well as the number of dimension-eight operators which will contribute to the process of interest $N_8$; for our figures we take $N_8=10$\footnote{A reasonable counting estimate would be to scale the 7 operators contributing at dimension-six by the ratio of number of operators at dimension-eight to dimension-six\cite{Henning:2015alf}; this yields $N_8\sim70$. However, it isn't clear what fraction of these operators will contribute with maximum energy growth.} throughout, but its impact is very minor as this study is dominated by statistical and systematic errors. Note that this $N_8$ parameter is the moral equivalent of the factor in scales explored when determining an error approximation in e.g. a QCD calculation. In the interest of not forcing the dimension-eight Wilson coefficients to be parametrically either large or small in comparison to the Wilson coefficients we are measuring at dimension-six, we choose
\begin{equation}
c_8=\frac{1}{2}\sum_k |C_k|,
\end{equation}
where $C_k$ is the Wilson coefficient of the dimension-six operator labeled by $k$; the factor of 1/2 amounts to an averaging, due to the presence of only two measurable linear combinations of Wilson coefficients at dimension-six. The two error distributions generated by the squared contributions of our exemplar operators $\cqq$ and $\cqu$ are shown in \cref{fig:sigdist} as error bars on the signal distribution. It is notable that these error distributions are larger in the more central region.

Throughout this article, all Monte-Carlo simulations are performed using \texttt{MadGraph5 v.2.6.2}~\cite{Alwall:2014hca} to generate three different distributions of the invariant dijet mass $\mjj$ at LO in QCD for the three terms in equation~(\ref{eq:consistency}). The first distribution corresponds to the SM, the second distribution to our signal prediction due to interference of the relevant SMEFT operators, and the last distribution corresponds to the theoretical error. Parton showering was performed by \texttt{Pythia 8}~\cite{Sjostrand:2014zea} and Detector simulation was provided by \texttt{Delphes v.3.4.1}~\cite{deFavereau:2013fsa}. The relevant SMEFT operators were implemented as a \texttt{FeynRules} model~\cite{Alloul:2013bka} supplied by~\cite{Brivio:2017btx,Aebischer:2017ugx}; we chose to utilize the $\alpha$-scheme for input parameters.

\section{Dijets at Tevatron}
\label{sec:tevdijet}

The Tevatron measurement which we recast to perform this analysis is a D0 publication of the dijet production cross section differential in both the dijet mass $m_{jj}$ and the absolute value of the rapidity of the more-forward jet $|y|_{\rm max}=\text{max}(|y_1|,|y_2|)$, where $y_i$ is the rapidity of the $i$th hardest jet~\cite{MandyRominsky:2009yya}. This measurement was made using $\lumiSeven$ of data. The measurement allowed for the presence of further hadronic activity, but only measured the properties of the two jets with highest transverse momentum $p_T$. The rapidity region covered ranges from $|y|_{\text{max}}=0$ to $|y|_{\text{max}}=2.4$. This rapidity region was split into six bins of equal size; within each maximum rapidity bin, a variable binning was used in $m_{jj}$ to ensure adequate statistics in each bin.

We utilized the signal and error spectra presented in \cref{fig:sigdist}, in conjunction with the D0 background calculation performed at NLO in QCD, to determine the sensitivity of this measurement to the SMEFT hypothesis for various Wilson coefficient values and cutoff scales. Due to its being based on a relatively small dataset, the D0 measurement has appreciable statistical errors on its bin contents. It also has systematic errors of comparable size. We combine these errors in quadrature with each other and with the theoretical error inherent in the SMEFT perturbation expansion to reach a value for the total error in each bin. Ultimately, these errors are too great to enable us to place a bound on the SMEFT, as will be discussed in greater detail in \cref{sec:results}.

We also explored the potential of a hypothetical analysis which utilized the full Tevatron dataset, rather than the small fraction of the data presented in this measurement, but this necessitated additional assumptions about the behavior of the systematic errors. There are two limiting cases which we consider here for these errors. The optimistic prediction hopefully assumes that the systematic errors will improve with increasing data in the same way that statistical errors do, because they are in large part derived from other measurements in control regions which should improve at that pace, and the pessimistic assumption treats these errors as a constant percentage error, imagining instead that the data considered in the D0 study was already enough to reach the true detector resolutions, and thus no further improvement could be expected on a fractional basis in these errors. The true case likely lies between these two extremes, of course.

\section{Results}
\label{sec:results}

All of our recast results are derived using a $\chi^2$ statistical treatment; as the systematic errors are asymmetric we utilize the error in the direction of the predicted deviation for that portion of the total error. We studied three different fitting strategies to achieve maximal sensitivity to the SMEFT signal in this analysis. Due to the difficulty of deriving a bound on the SMEFT contributions, we have focused here on the linear combination which gives a larger signal contribution, corresponding to the more central case from \cref{eq:centralDistr}.

\begin{figure}[t]
\includegraphics[width=.45\textwidth]{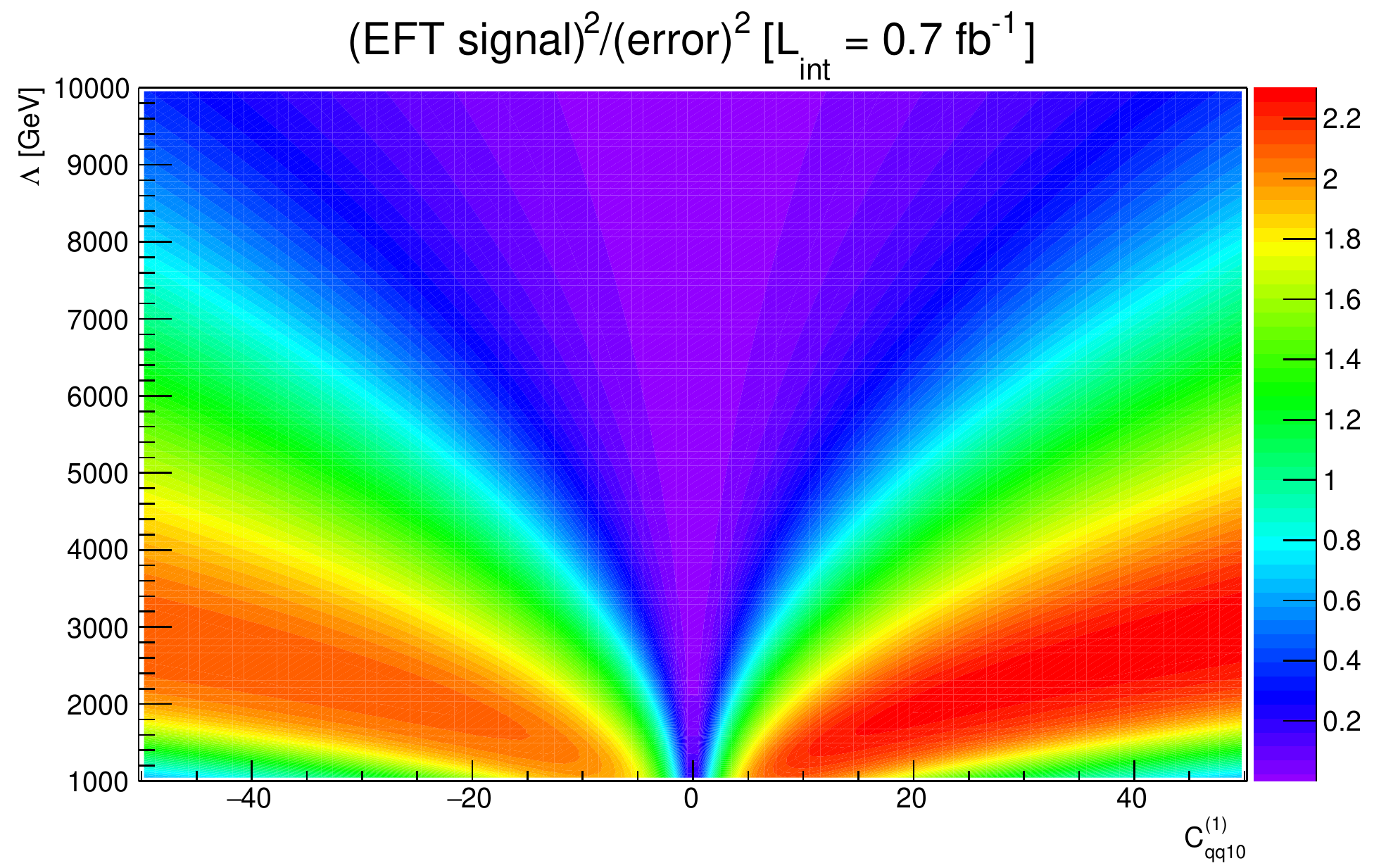}\hspace{0.1\textwidth}
\includegraphics[width=.45\textwidth]{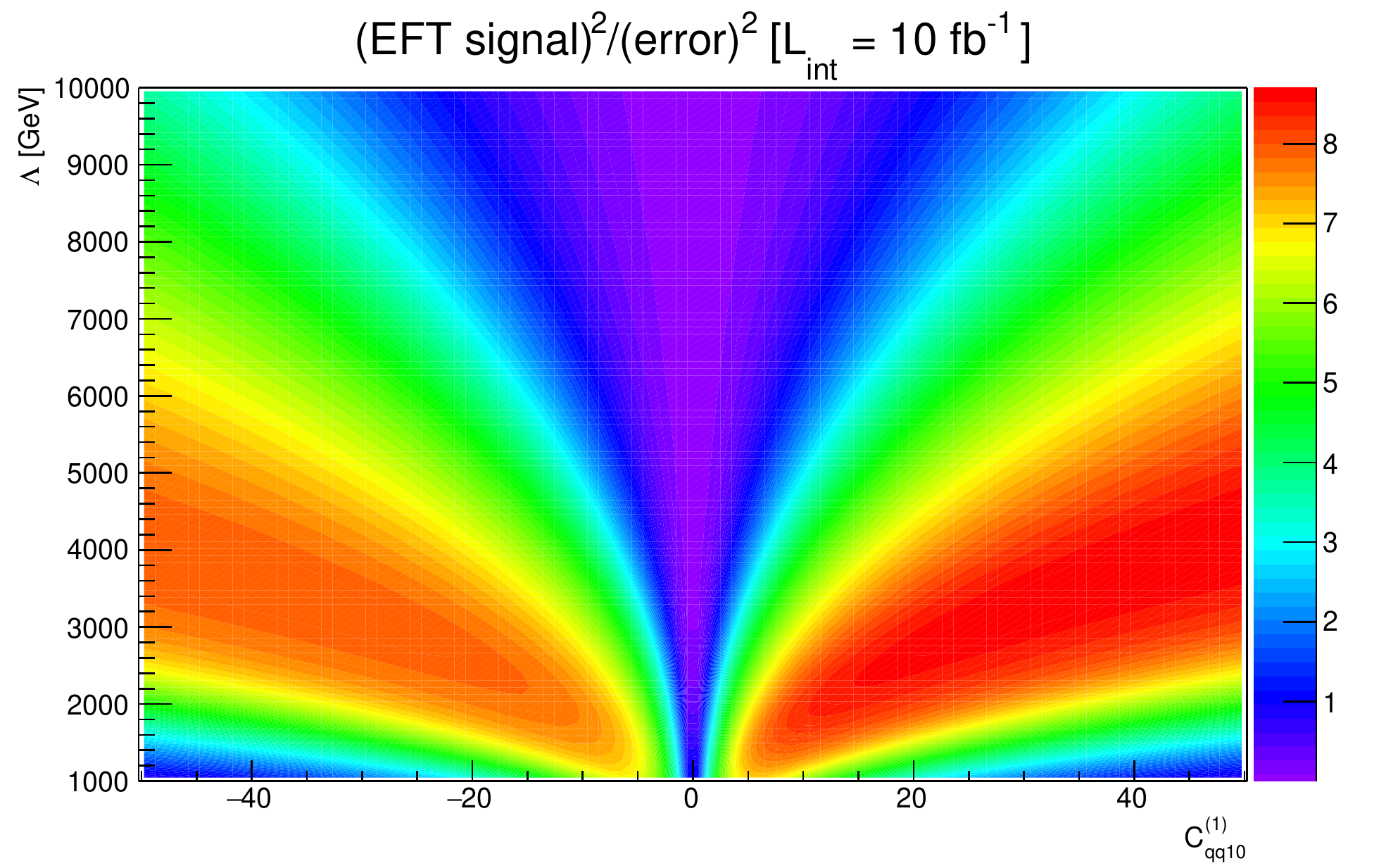}
\caption{\label{fig:results1}Full-spectrum fitting $\chi^2$ sensitivities to EFT contributions in dijet production in the current D0 analysis (left) and in an analysis using the full $\lumi$ Tevatron dataset (right). The full dataset sensitivity assumes that systematic errors will decrease in percentage terms similarly to statistical errors.}
\end{figure}

In our first analysis, we used the entire measured spectrum from D0, which has a total of 71 two-dimensional bins in $m_{jj}$ and $|y|_{\rm max}$. After determining the overall efficiency of the experimental reconstruction (which we don't trust Delphes to adequately reproduce in Monte Carlo) by fixing an efficiency for the SM background, we have 70 degrees of freedom in our fit. The predicted sensitivity to SMEFT effects in this case is presented in \cref{fig:results1} as a function of the cutoff scale $\Lambda$ and the Wilson coefficient $\cqq$, which contributes most to the signal. Note in particular the z-axis normalization; these values of $\chi^2$ are far from the critical value for a $2\sigma$ sensitivity, indicating that we do not expect the D0 search to have detected this effect. In the right-hand panel you see the scale-up for the most optimistic systematic error assumptions to an integrated luminosity of $\lumi$; even with the systematic errors decreasing as though they were purely statistical in nature, we remain far from sensitive in this analysis. We then investigated which bins were most sensitive, and selected those alone to fit, hoping to retain the majority of our $\chi^2$ value while reducing significantly the threshold for sensitivity. Even with this technique, though, our signal remained statistically undetectable.

As a second alternative approach we attempted a cut-and-count style analysis, where the bins of greatest sensitivity were combined into a single signal region, combining the data, background, and signal values to yield one effective bin. Each source of error required some thought to yield the correct combination however, and we ended up with two distinct treatments. The statistical error, of course, always combines in quadrature when summing bins, and the theoretical errors are prone to correlation, so we treated them as summing linearly rather than in quadrature. The difference in treatments again depends on what we believe about the systematic errors. Our more optimistic approach was to combine the systematic errors in quadrature as well, neglecting any correlations, yielding a total error for the signal region given by
\begin{align}\label{eq:treat1ofErrors}
\begin{aligned}
\sigma_\text{tot}^2=&  \sum_{i}\sigma^2_\text{stat,i}+\sum_{i}\sigma^2_\text{syst,i}+\left(\sum_{i}\sigma_\text{theo,i}\right)^2~,
\end{aligned}
\end{align}
where $i$ labels the $i$th bin that is part of the combined signal region. A more conservative alternate treatment which we explored presumes that the systematic errors are strongly correlated, and thus combine linearly rather than in quadrature.

\begin{figure}[t]
\includegraphics[width=.45\textwidth]{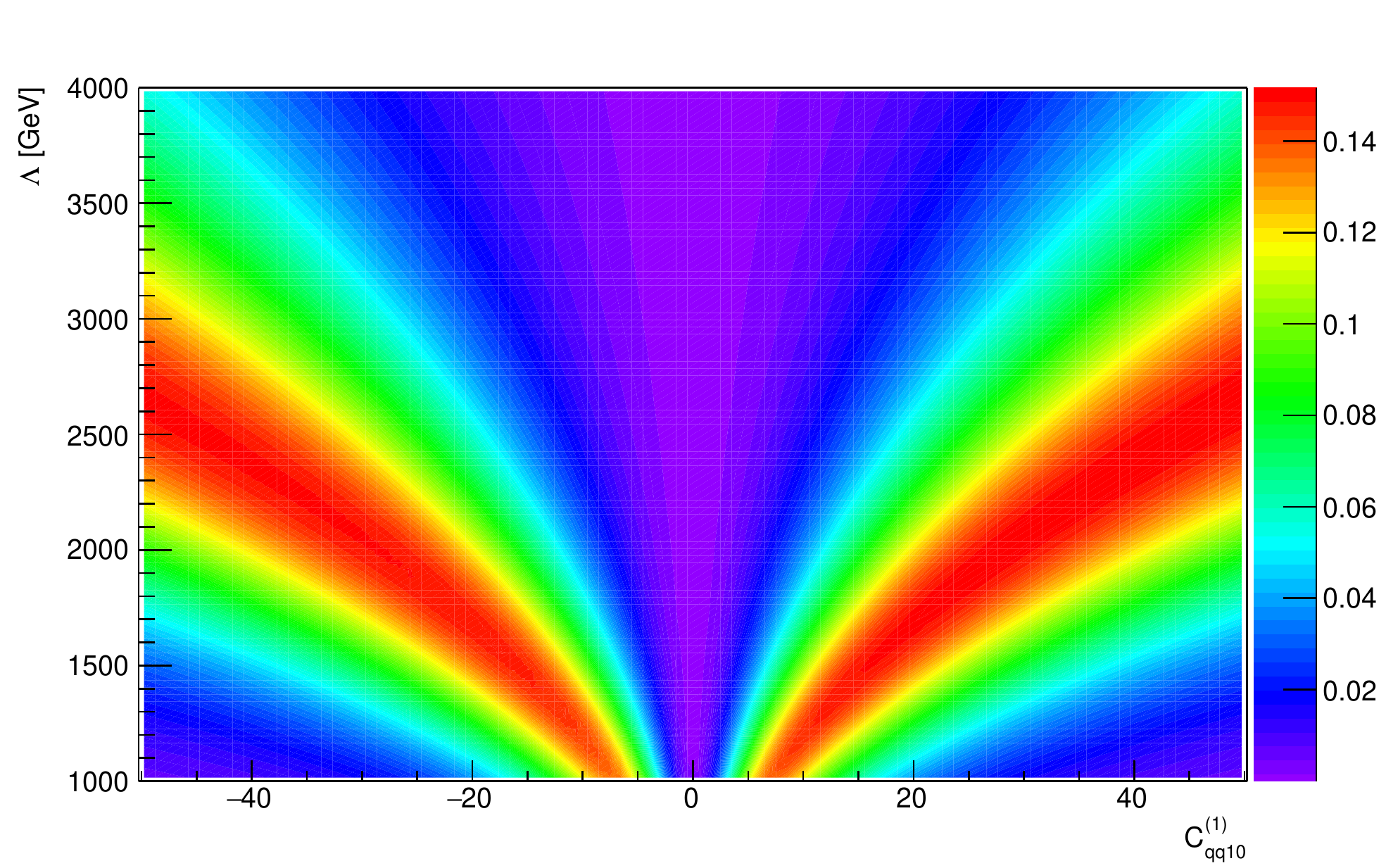}\hspace{0.1\textwidth}
\includegraphics[width=.45\textwidth]{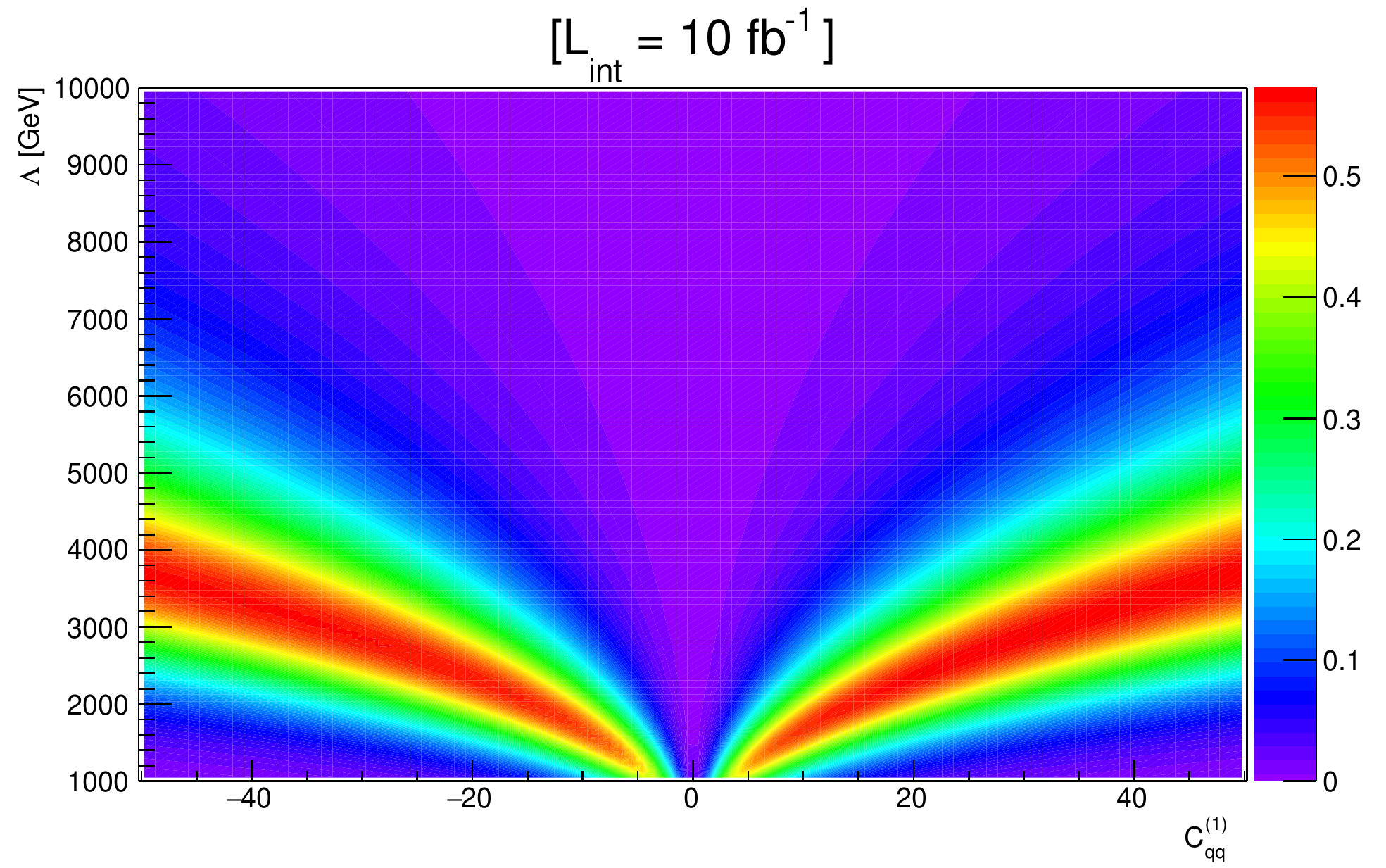}
\caption{\label{fig:results2}Cut-and-count sensitivity to EFT contributions in dijet production in the current D0 analysis (left) and in an analysis using the full $\lumi$ Tevatron dataset (right). The full dataset sensitivity assumes that systematic errors evolve with luminosity in the same way that statistical errors do, and both of the figures treat systematic errors as uncorrelated between bins.}
\end{figure}

The results of the more optimistic, uncorrelated systematic error treatment of a cut-and-count recast analysis are presented in \cref{fig:results2}. Here, because we are now considering only one bin, the color axis is trivially read as standard deviations, but the sensitivity remains far weaker than needed to claim any potential for D0 to measure or bound the SMEFT effect on dijet production. Again, the left panel presents the direct recast of the D0 measurement using $\lumiSeven$ of data, and the right panel presents the sensitivity possible using the full Tevatron dataset of $\lumi$. The regions of final-state kinematic space that have been combined to form the single signal bin are in the high-$m_{jj}$ and high-$|\eta|_{\rm max}$ areas, where the signal prediction is simultaneously relatively strong compared to the background (thanks to being at high energy) and relatively well-understood, with $\mathcal{O}\left(\frac{1}{\Lambda^4}\right)$ errors suppressed by their centrally-peaked structure at high rapidities.

\section{Conclusions}
\label{sec:conc}

We conclude that the Tevatron, and thus likely all previous hadron colliders as well, do not have the necessary precision in their dijet production measurements to be able to constrain consistently-calculated SMEFT effects on dijet production. This will leave the field of SMEFT parameterizations without any direct tree-level constraints on four-quark operators if the NP scale is below approximately 5 TeV. However, were the new physics scale that low for these interactions, we would expect to have seen the particle responsible for them at the LHC in at least a goodly fraction of cases. Nonetheless, this raises the potential specter of remaining flat directions in the parameter space of SMEFT Wilson coefficients for relatively low-scale NP.

The addition of further new observables to the list of those which could be used in a consistent global fit of the SMEFT will continue to ameliorate this problem by providing new linearly-independent constraints in the parameter space; it is plausible that these flat directions will ultimately be fully constrained by indirect measurements of the four-quark operators.

Different techniques for raising these flat directions in the absence of sufficient constraints at linear order in the Wilson coefficients have been proposed, but none of them avoid sensitivity to unknown effects of comparable magnitude from dimension-eight operators. Nonetheless, these flat directions are not by any means a death knell to the potential value of an EFT global fit; the tight correlation between Wilson coefficients of operators involving different fields necessary to exploit the flat directions is not a generic feature of UV models, and as such the fit will still provide interesting constraints. This also suggests an interesting avenue of model building, to explore what features of a UV model would successfully exploit one of these flat directions as a result of symmetry or other model-building constraints, rather than simply fine-tuning unrelated parameters in the UV.

\section*{Acknowledgments}
The authors acknowledge many helpful discussions with Stefan Baumgart throughout the development of this article. The work of WS was supported in part by the Alexander von Humboldt Foundation, in the framework of the Sofja Kovalevskaja Award 2016, endowed by the German Federal Ministry of Education and Research, and was partly performed at the Aspen Center for Physics, which is supported by National Science Foundation grant PHY-1607611.

\bibliography{FCNCbib}

\end{document}